\PassOptionsToPackage{table,xcdraw}{xcolor}
\documentclass[conference]{IEEEtran}
\IEEEoverridecommandlockouts
\usepackage{cite}
\usepackage{amsmath,amssymb,amsfonts}
\usepackage{multirow}
\usepackage{graphicx}
\usepackage{amsmath,amssymb,amsfonts}
\usepackage{algorithmic}
\usepackage{graphicx}
\usepackage{textcomp}
\usepackage{setspace}
\usepackage{xcolor}
\def\BibTeX{{\rm B\kern-.05em{\sc i\kern-.025em b}\kern-.08em
    T\kern-.1667em\lower.7ex\hbox{E}\kern-.125emX}}
\usepackage[utf8]{inputenc}
\usepackage[unicode=true, colorlinks = false]{hyperref}
\usepackage[sectionbib,sort]{natbib}
\setcitestyle{numbers}
\setcitestyle{square}
\usepackage{mathtools}
\usepackage{pdflscape}
\usepackage{afterpage}
\usepackage{url}
\usepackage{amssymb}
\usepackage{array}
\let\oldnl\nl
\usepackage{lscape}
\usepackage{yfonts}
\usepackage{enumitem}
\usepackage{amsfonts}
\usepackage{mathrsfs}
\usepackage{tikz}
\usepackage{mathdots}
\usepackage{booktabs}
\usetikzlibrary{fadings}
\usepackage{makecell}
\usetikzlibrary{patterns}
\usetikzlibrary{shadows.blur}
\usetikzlibrary{shapes}
\usepackage{csquotes}
\usepackage{tabularx}
\usepackage{multirow}
\usepackage{pgfplots}
\usepackage{pgf, tikz}
\usepackage{pgfplotstable}
\usepackage[linesnumbered,ruled]{algorithm2e}
\SetKwRepeat{Do}{do}{while}
\usepackage{lineno}
\usepackage{mathptmx}
\usepackage{graphicx}
\usepackage{cite}

\usepackage{amsmath}
\usepackage{float}
\definecolor{brightturquoise}{rgb}{0.03, 0.91, 0.87}
\definecolor{brightgreen}{rgb}{0.4, 1.0, 0.0}
\definecolor{candyapplered}{rgb}{1.0, 0.03, 0.0}
\definecolor{gold}{rgb}{0.83, 0.69, 0.22}
\let \oldnl \nl
\newcommand{\nonl}{\renewcommand{\nl}{\let \nl \oldnl}}

\DeclareMathAlphabet \mathbfcal{OMS}{cmsy}{b}{n}
\usetikzlibrary{arrows,positioning,calc,intersections,decorations.pathmorphing,positioning,decorations.pathreplacing,patterns,shapes.geometric}
\pgfplotsset{compat = 1.3}
\pgfdeclarelayer{background}
\pgfdeclarelayer{foreground}
\pgfsetlayers{background,main,foreground}
\usetikzlibrary{arrows, automata, trees}
\modulolinenumbers[5]
\DeclareFontFamily{OT1}{pzc}{}
\DeclareFontShape{OT1}{pzc}{m}{it}{<-> s * [1.10] pzcmi7t}{}
\DeclareMathAlphabet{\mathpzc}{OT1}{pzc}{m}{it}

\def \addlegendimage{\csname pgfplots@addlegendimage\endcsname}
\usepackage{orcidlink}
\usepackage{multirow, booktabs, siunitx, caption}

\begin{document}

\author{\IEEEauthorblockN{Adel Albshri\IEEEauthorrefmark{1}\IEEEauthorrefmark{2}\orcidlink{0000-0001-9107-2429}, Ali Alzubaidi\IEEEauthorrefmark{3}\orcidlink{0000-0002-9653-4474} and Ellis Solaiman\IEEEauthorrefmark{4}\orcidlink{0000-0002-8346-7962}}

\IEEEauthorblockA{\IEEEauthorrefmark{1}Newcastle University, School of Computing, UK, Email: a.albshri2@ncl.ac.uk}

\IEEEauthorblockA{\IEEEauthorrefmark{2}University of Jeddah, College of Computer Science and Engineering, Saudi Arabia, 
Email: amalbeshri@uj.edu.sa}

\IEEEauthorblockA{\IEEEauthorrefmark{3}Umm Al-Qura University, College of Computing, Saudi Arabia, Email: aakzubaidi@uqu.edu.sa}

\IEEEauthorblockA{\IEEEauthorrefmark{4}Newcastle University, School of Computing, UK, Email: ellis.solaiman@ncl.ac.uk}}

\title{A Model-Based Machine Learning Approach for Assessing the Performance of Blockchain Applications}
\maketitle
\thispagestyle{empty}
\pagestyle{empty}

\begin{abstract}
The recent advancement of Blockchain technology consolidates its status as a viable alternative for various domains. However, evaluating the performance of blockchain applications can be challenging due to the underlying infrastructure's complexity and distributed nature. Therefore, a reliable modelling approach is needed to boost Blockchain-based applications' development and evaluation. While simulation-based solutions have been researched, machine learning (ML) model-based techniques are rarely discussed in conjunction with evaluating blockchain application performance. Our novel research makes use of two ML model-based methods. Firstly, we train a $k$ nearest neighbour ($k$NN) and support vector machine (SVM) to predict blockchain performance using predetermined configuration parameters. Secondly, we employ the salp swarm optimization (SO) ML model which enables the investigation of optimal blockchain configurations for achieving the required performance level. We use rough set theory to enhance SO, hereafter called ISO, which we demonstrate to prove achieving an accurate recommendation of optimal parameter configurations; despite uncertainty. Finally, statistical comparisons indicate that our models have a competitive edge. The $k$NN model outperforms SVM by 5\% and the ISO also demonstrates a reduction of 4\% inaccuracy deviation compared to regular SO.
\end{abstract}


\section{Introduction}
\label{Introduction}

Blockchain technology, a form of \textit{Distributed Ledger Technology} (DLT), has seen increasing adoption across various sectors, including healthcare, supply chain management, and the Internet of Things (IoT). This is due to its decentralized nature, resistance to tampering, and features such as consistency, anonymity, and traceability. These features make it an excellent choice for applications requiring high levels of security and accountability \cite{Zhang2023} \cite{albshri2022investigating}.

Nevertheless, configuring a blockchain optimally can pose challenges, as applications' requirements vary significantly. Factors influencing these requirements include the nature of stored data, the frequency and concurrency of transactions, the number of validating nodes, and infrastructure specifications such as CPU, memory, network bandwidth, and Input/Output speed. The constraints that these applications must account for can further complicate the development process, thus affecting the overall performance of the blockchain-based application in terms of throughput, latency, and the rate of successful/failed transactions \cite{fan2020performance}.

This study draws inspiration from a hypothetical scenario where a healthcare organization is contemplating the integration of blockchain technology into its operations. To gauge the feasibility and potential success of this project, certain performance metrics, such as transaction volume, average transaction time, and transaction success rate, amongst others, can be utilized. These metrics can be leveraged for one of specific purposes outlined subsequently:

\begin{enumerate}
    \item Predicting how the blockchain-based application will perform under certain conditions and preset configuration parameters, given the limitation of the available resources. 
    
    \item Vice versa, given a target performance level, the task is to estimate the right configuration parameters. This is to answer questions like what configurations should be in place to enable an IoT-enabled hospital to achieve a blockchain throughput of at least 1000 transactions per second.
\end{enumerate}

To approach the selection and implementation of blockchain-based applications systematically, it is necessary to conduct a comprehensive evaluation of the application's requirements. Numerous simulation frameworks for this purpose have been proposed \cite{AlbshriSimulators}. However, due to blockchain systems' complexity, it is challenging to provide a comprehensive and accurate representation of a specific blockchain application's performance. The interdependency and wide range of parameters in a blockchain system make achieving an accurate performance evaluation a significant challenge. 

On another front, machine learning (ML), a subfield of artificial intelligence, uses historical data to develop algorithms and statistical models that aim for optimal performance \cite{guo2021using}. This paper mainly focuses on the supervised learning approach, specifically on classification and optimization algorithms.

In the realm of machine learning, classification is concerned with understanding and sorting data into predetermined groups or ``sub-populations". Classification algorithms use labelled training data to determine whether an object belongs to a predefined category. They identify recurring patterns and common features, thereby enabling "pattern recognition". The efficiency of these algorithms is evaluated based on their ability to classify objects correctly. In this study, we focus on two well-known classification algorithms: the $k$ nearest neighbour ($k$NN) algorithm \cite{Hamed2022} \cite{Hamed2021}, and the support vector machine (SVM) algorithm \cite{Panigrahi2023}.

Swarm optimization, a machine learning technique, is gaining attention due to its ability to efficiently find near-optimal solutions for complex problems, even with limited resources, such as processing power or time, and incomplete or imprecise knowledge about the problem \cite{Rodriguez2023} \cite{salem2022effective}. To overcome the limitations of traditional optimization methods, several ML algorithms have been proposed, such as Harris Hawk Optimization (HHO) \cite{Hfeature},  Grey Wolf Optimization (GWO) \cite{Hamed2021efficient}, Artificial Bee Colony (ABC) \cite{Handur2023artificial}, Ant Colony Optimization (ACO) \cite{Karimi2023semiaco}, Particle Swarm Optimization (PSO) \cite{Ganesh2023}, and Salp Optimization (SO) \cite{Harish2023energy}. These algorithms provide robust optimization capabilities.

Another interesting machine learning algorithm is the Rough Set Theory (RST) \cite{hamed2021distributed}, which provides a formal approach to approximate conventional or crisp sets using lower and upper approximations. If the lower and upper approximations are identical, RST provides a crisp set. If the approximations are different, variations of RST may result in rough sets.

This study aims to use ML techniques to help mitigate some of these challenges by providing a more comprehensive and accurate evaluation of blockchain performance. The main contributions of this paper are:

\begin{enumerate}
    \item An ML model that utilizes the $k$NN algorithm to predict a blockchain system's performance, considering various parameters such as the number of nodes, number of miners, and number of transactions.
    \item An improved Salp Optimization (ISO) algorithm leverages the capability of a rough set in dealing with uncertainties to predict optimal parameter configurations for a given metric value.
\end{enumerate}

The remainder of this paper is structured as follows. Section \ref{related} provides a brief review of the related works in the literature. Section \ref{model} proposes our two models for predicting the overall blockchain performance and estimating the optimal configuration parameters, respectively. Section \ref{experiments} conducts several experiments to validate the proposed models. Finally, Section \ref{conc} concludes the paper and discusses potential future work.

    

\section{Related Work}
\label{related}

In the quest to systematically and strategically evaluate blockchain performance, a variety of approaches have been proposed. Some of these methodologies centre around performance analysis via monitoring and observation of blockchain networks. For instance, a log-based blockchain monitoring framework was put forward by Zheng et al. \cite{zheng2018detailed}. Although this form of monitoring aids in identifying recurring patterns and observing performance, it lacks an intuitive and straightforward mechanism to propose the optimal configuration for peak performance. 

To address this, several studies in the literature propose benchmarking blockchain networks by applying synthetic workloads in a controlled evaluation environment. Dinh et al. \cite{dinh2017blockbench}, for example, introduced a benchmarking framework, named BlockBench, for evaluating and analysing the performance of private blockchain platforms, such as Hyperledger Fabric and a private version of Ethereum. This was conducted with a focus on key aspects including latency, throughput, fault tolerance, and scalability \cite{salem2022fuzzy}. Hyperledger Caliper \cite{Caliper} is another such tool, aimed at gauging the performance of various Hyperledger blockchains, namely Fabric, Sawtooth, Burrow, and Ethereum. This tool evaluates performance based on four critical metrics: throughput, latency, transaction success rate, and resource utilization. Despite the advantages of benchmarking in ensuring precise evaluation and measurement, the process remains primarily a trial-and-error endeavour that does not guarantee an automated discovery of optimal performance. 

Several studies, in their pursuit of optimal performance, have focused on characterizing performance features of existing blockchain platforms under varying workloads and supported consensus algorithms. By doing so, they aim to reveal the maximum attainable performance in terms of throughput and latency characteristics. For instance, a comprehensive performance analysis of Ethereum was conducted by Rouhani and Deters \cite{bez2019scalability}, wherein they assessed the two most widely used Ethereum clients - the Proof-of-Work (PoW)-based Geth and the Proof-of-Authority (PoA)-based Parity. Similarly, an in-depth analysis of the Quorum blockchain's performance was performed by Baliga et al. \cite{baliga2018performance}. These studies have provided insights into performance approximations under a given set of conditions. 

To circumvent the challenges associated with real blockchain deployment, several blockchain simulation frameworks have been proposed. The aim here is to facilitate the investigation of various configuration parameters' impact on overall performance for different scenarios, requiring the least possible effort. For example, the frameworks proposed by Alharby and Moorsel \cite{alharby2020blocksim} and Pandey et al. \cite{Pandey2019} enable the discrete-event dynamic simulation of blockchain platforms using various configurations to assess overall performance. Other blockchain simulation efforts are covered in \cite{AlbshriSimulators}.

Despite the facilitation offered by simulation, it shares a common problem with real blockchain benchmarking – the reliance on predetermined parameters for modelling blockchain system behaviour. This approach could fall short of achieving the best possible performance due to the difficulty in determining the optimal values for configuration parameters. Hence, this work investigates the use of machine learning techniques to estimate blockchain performances and suggest optimal configuration parameters, thereby aiming to achieve the best possible performance.
\section{Proposed Models}
\label{model}

\subsection{Preliminaries}
Assume a number of configuration parameters (input) and performance metrics (output) of a blockchain-based solution as follows:
\begin{enumerate}
    \item Configuration \textit{Parameters}, $P$: the set of $l$ parameters $P = \{p_1,p_2,\ldots,p_l\}$ represents the \textit{input configuration} of the blockchain network such as the quantity of participating nodes, transactions frequency, payload size, selected consensus mechanism, and so forth.
    
    \item Performance \textit{Metrics}, $M$: the set of $n$ metrics $M = \{m_1,m_2,\ldots,m_n\}$ represent the \textit{conditional outputs} with respect to the given parameters $P$ such as network throughput and latency.
\end{enumerate}
Hypothetically, there is a strong correlation between configuration parameters and produced metrics. Therefore, we investigate the following: 

\begin{itemize}

    \item Employing $k$NN algorithm as a regression tool for predicting the overall performance in terms of each metric $m \in M$ of a blockchain-based application based on a given set of configuration parameters $P$.
    \item Employing the Salp Swarm Optimization (SO) algorithm to determine the optimal configuration parameters $P$ based on a target level for each performance metric $m \in M$ 
\end{itemize}

\subsection{$k$NN Regression Algorithms for Performance Predication}
\label{knn}
To identify commonalities, $k$NN algorithms compare a given set of parameters ($P_0$) with unknown values of performance metrics to their $k$ neighbours. The commonalities are usually computed using a distance measure. The idea is that the set of parameters $P_0$ will be closer to the set of parameters $P_i$ of similar characteristics. $k$NN trains vectors with class labels in a multidimensional feature space. Each training data row has its parameters setup and decision values. Here, measurements are decision-conditional features. Only the algorithm's training samples' feature vectors and class labels are stored. Averaging the metric values of $k$ nearby objects should yield the anticipated value. Given a dataset $D$ with $l$ features (configuration parameters) and $m$ performance metric, we refer to the parameter value $j$ of object $i$ as $v_{i,j}$. For example, $v_{2,5} = 6$ means that parameter number 2 of object number 2 has value 6. Moreover, the decision value $m_k$ of object $i$ is referred to as $v_{i,d}$.

$k$NN depends heavily on a distance measure. Euclidean distance is a typical distance metric for continuous values (parameters). The Euclidean distance $l\left(u_{0},u_{i}\right) $ between two different
objects, $u_{0}$ and $u_{i}$ is given by%
\begin{equation}
	l\left( u_{0},u_{i}\right) =\sqrt{\left( \mathbf{V}_{u_{0}}^{\prime }-%
		\mathbf{V}_{u_{i}}^{\prime }\right) ^{T}\left( \mathbf{V}_{u_{0}}^{\prime }-%
		\mathbf{V}_{u_{i}}^{\prime }\right) }\text{,}  \label{EucMat}
\end{equation}%
where%
\begin{equation*}
	\mathbf{V}_{u_{k}}^{\prime }=<v_{u_{i},a_{1}^{\prime }},v_{u_{i},a_{^{\prime
			}2}},\ldots ,v_{u_{i},a_{m}^{\prime }}>\text{,}
\end{equation*}%
The proposed algorithm employing the previous steps is shown in Algorithm \ref{knnAlgo}.
\begin{algorithm*}[ht]
    \scriptsize
	\SetKwInOut{Input}{Input}
	\SetKwInOut{Output}{Output
 }
	\Input{$D$ //Training data\newline
	$u_0$ //Unknown query object\newline
	$k$ //Number of nearest neighbour
	}
	
	\Output{$M$ //Set of predicted metrics}
	$L:=\emptyset$
	
	\For{\textbf{each} object $u_i \in D$ }{
	
	Compute the Euclidean distance between $u_0$ and $u_i$ as per Eq. (\ref{EucMat}) and add it to $L[i]$
	}

        Sort $L$ in ascending order
        
	Find the first $k$ objects in $L[i]$ with the least distance value
	
	\For{\textbf{each} metric $m_i \in M$}{
	
	Compute the value of $m_i$ for the unknown object $u_0$ by averaging the corresponding metric values of the $k$ neighbouring  objects

       \nonl $m_j:= \frac{1}{k}\sum_{j=1}^k v_{j,d}$, where $v_{j,d}$ is the decision value of object $u_j$ in the first $k$ objects in $L$
        
	}
	
	\caption{$k$NN regression algorithm for blockchain metircs prediction}\label{knnAlgo}
\end{algorithm*}

\subsection{Improved Salp optimization (ISO) algorithm}
\label{sec:SO}

Each salp has a number $i$, where $i=1,2,\ldots,\mathscr{P}$ and an identifier indicating whether it is a leader or not. The numerals are permanent, but the identifiers may vary between iterations. In the initial iteration, the $P$ salps occupy arbitrary ``positions" in the chain, i.e., they simply adhere to the chain. A ``position" is a location vector that describes a set of parameter values. The algorithm finalizes the iteration by identifying the $m$ salps with the greatest performance as leaders, moving them to the front of the chain, and allowing them to share their position data (location vectors) with the non-leaders. In other words, the algorithm accomplishes the parameter values identification assignment in two successive steps: the exploration step and the exploitation step, each of which is described in greater detail below.

\textbf{ISO exploration step}

Salp $i\geq 1$ has in iteration $k\geq 1$ a location vector $\mathbf{S}_{i_{k}}=%
\left[ s_{1},s_{2},\ldots,s_{n}\right]$, the values of $y_{j}$ has different ranges. Therefore, we feed the algorithm by the separate range of each $s_j$. In subsequent cycles, this $s_j$ is constantly updated. The parameter vector defines the parameters' values. The $s_j$ specifically reflects the value of parameter $j$.For example, $\mathbf{S}_{3_{2}}=\left[ 4,2,1,0.064%
\right] $ means that salp $3$ in iteration $2$ is representing parameter configuration for four parameters with values 4,2,1,0.064, respectively.

At iteration $1$, each salp is started by an \emph{randomly} generated parameter vector of the $l$ original parameters, acquiring an initial parameter vector. Remember that the random values are chosen with the $s_j$ bounds in mind. This parameter vector is changed on each cycle. The dependence function evaluates the fitness of the parameter, and vectors, and also acts as an ambiguity-relaxing tool. 

Specifically, the dependency value $\gamma $ is computed by the end of each iteration $k\geq 1$,  for the parameter vectors of all $P$ salps in the chain is calculated. The salps with the greatest $\gamma $ values are then designated as leaders. Those in charge are said to be closer to the ideal parameter setting than the others.

\textbf{ISO exploitation steps}

Let the set of $p$ leaders in iteration $k\geq 1$ be $\mathbb{P}_{_{k}}$. A non-leader salp $i$ gets its updated parameter vector $\mathbf{S}_{i_{k+1}}$ in two stages in the subsequent iteration $k+1$ of the algorithm. Each leader salp modifies its parameter vector in the first phase in the manner described below.

\begin{equation}
    \mathbf{S}_{i_{k}} = \mathbf{S}_{i_{k - 1}} + r^2 (ub - lb)  + r lb
\end{equation}

Similarly, each non-leader salp $i$, $i\notin \mathbb{P%
}_{_{k}}$, will calculate mean difference of $p$ vectors (one for each $j\in \mathbb{P}_{_{k}}$) as follows. 
\begin{equation}
\mathbf{D}_{i,j}= \frac{1}{m}\left[\sum r_{1}\mathbf{S}_{i_{k}}-\mathbf{S}_{j_{k}}\text{,\quad }%
j\in \mathbb{P}_{_{k}}\right]\text{,}  \label{Diff2}
\end{equation}%
where $r$ is given by
\begin{equation*}
    r = 2 e^{-\left(\frac{4m}{L}\right)}.
\end{equation*}

To this end, given a set of $p$ salps $\mathbf{S}_{i_{j}}$ at iteration $j$, some of the salps parameter vectors may have ambiguous values. The ambiguous values are those that do not lead to promising solutions. Therefore, it will be a hard task to update such vectors. This problem may worsen by getting trapped in the local minima. Consequently, we introduce a goodness function $\gamma$ depending on the well-known mathematical theory: rough set theory (RST), to solve such an issue. RST is known for its promising abilities in dealing with ambiguity through computing the approximation space, which is a set of approximations referred to as lower and upper. The former represents the set of objects with no ambiguity, while the latter represents the set of ambiguous objects. Assume we are optimizing the metric value $m_k$; i.e. $m_k$ is the input value. First, we compute the fitness value $f_{\mathbf{S}_{i_{j}}}$ of each salp $\mathbf{S}_{i_{j}}$ by computing the regression value using $k$NN algorithm as per Section \ref{knn}. Second, let $\tau$ be a user-defined value that serves as a threshold. The set of salps $\mathbb{S}^+_j$ at iteration $j$ having $f_{\mathbf{S}_{i_{j}}}>\tau$ are considered good; otherwise; the set $\mathbb{S}^-_j$ are considered ambiguous. 

\textbf{Definition 1 (Goodness function, $\gamma$):} Given finite set of $n$ salps $\mathbf{S}_{i_{j}}$, we compute the equivalence relation $E$ of each salp as follows.
\begin{equation}
    E_{\mathbf{S}_{i_{j}}} = \{\mathbf{S}_{k_{j}}| l(\mathbf{S}_{i_{j}}, \mathbf{S}_{k_{j}}) < \frac{1}{2} (|\mathbf{D}_{i,j} - \mathbf{D}_{k,j}|)\}
\end{equation}

The lower, upper and boundary approximations are given as follows.

\begin{equation}
\underline{Apr}(\mathbb{S}^+_j) = \{ E_{\mathbf{S}_{i_{j}}}| : E_{\mathbf{S}_{i_{j}}} \subseteq \mathbb{S}^+_j  \}\
\label{eq:lw}
\end{equation}
\begin{equation}
\overline{Apr}(\mathbb{S}^+_j) = \{ E_{\mathbf{S}_{i_{j}}}| : E_{\mathbf{S}_{i_{j}}}| \bigcap \mathbb{S}^+_j  \neq \emptyset \}\
\label{eq:}
\end{equation}
\begin{equation}
BND(\mathbb{S}^+_j) = \overline{Apr}(\mathbb{S}^+_j) - \underline{Apr}(\mathbb{S}^+_j) 
\label{bnd}
\end{equation}
Finally, the goodness of the upper approximation is given by
\begin{equation}
    \gamma = \frac{|\underline{Apr}(\mathbb{S}^+_j)|}{n}.
\end{equation}
$\blacksquare$

\begin{algorithm*}[ht]
\scriptsize
\linespread{0.2}
	\SetKwInOut{Input}{Input}
	\SetKwInOut{Output}{Output}
	\Input{$m$ //Metric value to be optimized\newline
		$P$ //Number of salps\newline
		$R$ //Number of iterations ($R \geq 2$)
	}
	
	\Output{$P$ //Set of $l$ parameters}

	\nonl //Initialization step:\\
	$k= 1$\\
	$\Gamma = [\ ]$ //An empty list to save the dependency of all salps\\
	\For{$i =1 $ \textbf{to} $P$}{
		Construct parameter vector $\mathbf{S}_{i_{k}}=\left[ s_{1},s_{2},...,s_{n}\right]$, 
		where $y_{j}$ is set randomly according to the parameter constraints.\\
		Calculate the fitness $f_{\mathbf{S}_{i_{k}}}$ using $k$NN model as per Algorithm \ref{knnAlgo}\\
		
	}
	Compute the goodness of the $P$ salps as per Definition 1\\
	Delete the salps appearing in $BND(\mathbb{S}^+_j)$ computed using Eq. (\ref{bnd})\\
	
	Construct set $\mathbb{L}_{_{k}}=\{i_1,i_2,,...,i_{m}\}$, where the $i_j$ are the indices of the highest $p$ values in $\Gamma$. //Tag top performing	salps as leaders.\\
	Regenerate the deleted salps with respect to the leaders\\
	\nonl //Iteration steps:\\
	
	\Do{($k < R$)}
	{
		$k = k + 1$\\
		$\Gamma = [\ ]$\\
		\For{$i = 1 $ \textbf{to} $P$}{
			\eIf{$i \in \mathbb{L}_{_{k-1}}$}{
				\nonl	//If salp $i$ is tagged as leader\\
				Calculate $f_{\mathbf{S}_{i_{k}}}$, using $k$NN and append it to $\Gamma$.\\
			}
			{
				\nonl //If salp $i$ is not tagged as leader\\
				Calculate parameter vector $\textbf{S}_{i_k}$ from $\textbf{S}_{i_{k-1}}$,
				as per (\ref{Diff2}). //Update parameter vector.\\
				Compute the goodness of the $P$ salps as per Definition 1\\
	            Delete the salps appearing in $BND(\mathbb{S}^+_j)$ computed using Eq. (\ref{bnd})\\
	          
			}
		}
		Construct set $\mathbb{L}_{_{k}}=\{i_1,i_2,...,i_m\}$, where the $i_j$ are the indices of the highest $p$ values in $\Gamma$. //Tag top performing salps as leaders.\\
		
		Assign the highest $f_{\mathbf{S}_{i_{k}}}$to $Hfit$.
		
		Set $P$ to the best salp\\
	}
	
	\caption{Improved Salp Optimization Algorithm (ISO)} \label{algo:SO}
\end{algorithm*}

Having said this, to improve ISO algorithm convergence and to avoid getting trapped in the local optima, the set of salps in the boundary region $BND(\mathbb{S}^+_j)$ is completely deleted are regenerated concerning the salps having high goodness values.

The processes mentioned above are used by the ISO pseudocode displayed in Algorithm \ref{algo:SO}. Only the first iteration, where $k=1$, uses the algorithm's initialization process. Then it executes a loop where a different method is used for every $k>1$ iteration. The computational cost of ISO may be calculated by using Algorithm \ref{knnAlgo} and noting that $P$ is the number of salps and $R$ is the number of iterations. The \emph{exploration} step involves ISO spanning $P$ parameter vectors. With $N$ salps in hands, computing the fitness function for each vector costs $O(N)$. The exploration phase thus costs $O(MN)$. Second, the ISO method changes each parameter vector $R$ twice at most during the \emph{exploitation} stage. As a result, this step's cost is $O(RMN2)$. The entire computing cost of the ISO method is $O(RMN2)$ since the exploitation step is the most important one.

\section{Experimental work}
\label{experiments} 
The proposed models were implemented using Python and executed on a system equipped with CentOS 7, a 2.4 GHz Intel Core i7 processor, and 16 GB of RAM. The code is available on GitHub\footnote{\url{https://github.com/AlbshriAdel/BlockchainPerformanceML}}. We conducted several experiments using the collected data, with two main objectives in mind. First, we aimed to test the $k$NN model's ability to predict blockchain performance accurately. Second, we aimed to test the ISO algorithm's ability to identify the best parameter configurations required to achieve a user-defined value for a specific metric, such as throughput.

\subsection{Data collection}
\label{dc}
There is currently no readily accessible public dataset tailored to the tasks outlined in this work. Furthermore, considering our objective to validate the proposed concepts, we elected to utilize a dataset derived from a simulation environment. This approach allows us to control the parameters involved and generate a diverse array of performance data.

In the simulation scenario used for our study, we employed the Raft consensus algorithm. As per the operational constraints imposed by Raft, we were compelled to operate with a single miner node. This constraint is inherent to the design of the Raft consensus protocol and is not a limitation of our study per se.

It is important to note that the training of machine learning models necessitates a substantial volume of historical data. Therefore, we generated the requisite data using a blockchain simulator. The specifics of the parameters ($P_i$) that we manipulated to alter the blockchain's characteristics are described in Table \ref{parameterDescription}. Our data generation approach provided us with the flexibility to adjust these parameters and collect a comprehensive dataset for our machine learning models.

\begin{table}[tbh]
\centering
 \caption{The description of the nine used parameters with their abbreviation, lower $L$ and upper $U$ bound of each.}
\label{parameterDescription}
\resizebox{\columnwidth}{!}{%
\footnotesize
\renewcommand{\arraystretch}{1.8}
\begin{tabular}{p{2.3cm}cp{3.4cm}ccc}
\hline

\multicolumn{1}{p{2.3cm}}{Parameter} & \multicolumn{1}{c}{Abb.} & \multicolumn{1}{p{3.4cm}}{Desc.}& \multicolumn{1}{c}{{Format}} &  \multicolumn{1}{c}{$L$} &  \multicolumn{1}{c}{$U$} \\ \hline

Number of nodes & $P_1$& The number of nodes participating in the blockchain network  & Integer& 3& 15\\ \hline

Number of miners& $P_2$& The number of miners participating in the blockchain network & Integer& 1 & 1\\ \hline

Consensus algorithm & $P_3$& Consensus Algorithm ``Raft" & String& -& - \\ \hline
\#transactions/ second & $P_4$ & The total number of transactions generated & Integer& 9& 1650\\ \hline
Max block size& $P_5$& The maximum amount of block size& Decimal& 1& 1\\ \hline
Max transaction size & $P_6$& The maximum transaction data size& Decimal& 0.064 & 0.064 \\ \hline
Min transaction size & $P_7$& The minimum transaction data size& Decimal& 0.001& 0.001\\ \hline
Block interval& $P_8$& Block processing time& Decimal& 0.05& 0.0099 \\ \hline
Simulation time& $P_9$ & The time taken for executing& Decimal& 1& 1\\ \hline
\end{tabular}
}
\end{table}

The simulated blockchain model is executed several times using different configuration values for the parameters described in Table \ref{parameterDescription}. During these runs, we thoroughly examined the data. Having identified the set of conditional features, we now turn our attention to the decision features, which include the set of performance metrics $M$. These features are computed based on conditional features and can be used to evaluate the performance of the blockchain. Table \ref{metricDescription} provides details about the metrics we have used.

\begin{table}[tbh]
    \centering
     \caption{The description of the thirteen used metrics with their abbreviation.}
     \label{metricDescription}
\resizebox{\columnwidth}{!}{%
\footnotesize
\renewcommand{\arraystretch}{1.8}
\begin{tabular}{p{2.3cm}cp{3.4cm}c}
\hline
 
\multicolumn{1}{p{2.3cm}}{{Metric}}& {Abb.} & \multicolumn{1}{p{3.4cm}}{{Desc.}} & {Format} \\ \hline

Total number of blocks & $M_1$& The number of blocks generated & Integer\\ \hline

Total number of blocks including transactions & $M_2$ & The number of blocks that contains transactions & Integer \\ \hline

Total number of transactions & $M_3$ & The number of transactions generated & Integer  \\ \hline

Total number of pending transactions & $M_4$ & The number of transactions not processed & Integer \\ \hline

Total number of blocks without transactions & $M_5$ & The number of empty blocks & Integer \\ \hline

Average block size & $M_6$ & The average blocks size & Decimal \\ \hline

Average number of transactions per block & $M_7$ & Average transactions per block & Decimal \\ \hline

Average transaction inclusion time & $M_8$& Average transaction time & Decimal \\ \hline

Average transaction size& $M_9$ & The average size of the transactions & Decimal \\ \hline

Average block propagation & $M_{10}$ & Average block time & Decimal \\ \hline

Average transaction latency & $M_{11}$ & The average time between transaction submission and confirmation & Decimal \\ \hline

Transactions execution & $M_{12}$ & Average number of transactions per block & Decimal \\ \hline

Transaction Throughput & $M_{13}$ & The rate of transactions throughput & Decimal \\ \hline

\end{tabular}
}
\end{table}

Note that computing the decision characteristics presented in Table \ref{metricDescription} in simulation mode requires computing the prior features, as shown in Table \ref{parameterDescription}. To count these features, we need to have access to the details of each block, which can be a time-consuming process. Therefore, we can define the issue as follows: we will use the ML method ($k$NN) and conditional features to directly forecast metric choice feature values. In the following sections, we will train the ML model to predict decision feature values using conditional features.

It is crucial to examine the statistical properties of the collected data to ensure the reliability of the subsequent ML results. Upon reviewing Table \ref{parameterDescription}, we observe that there are six numerical features ($P_5$, $P_6$, $\ldots P_9$) present in the dataset.

Prompted by this observation, we sought to gain insights into the dispersion and distribution of these numerical features. We specifically calculated the mean and standard deviation for these features to evaluate the skewness, or asymmetry, of the distribution in the dataset. Additionally, we conducted an examination for any missing values that might affect the analysis. 

The results of this comprehensive statistical analysis are detailed in Table \ref{statanalysis}. These preliminary findings will aid us in understanding the inherent characteristics of our dataset, thereby assisting in the formulation of more accurate machine learning models and predictions.
\begin{table}[tbh]
    \centering
    \caption{Statistical analysis (mean, standard deviation, std, minimum and maximum values) for numerical features (5 parameters and 8 metrics).}
    \label{statanalysis}
    \footnotesize
\renewcommand{\arraystretch}{1.8}
\begin{tabular}{ccccc}
\hline 
\multicolumn{1}{c}{Feature}
& \multicolumn{1}{c}{Mean}
& \multicolumn{1}{c}{Std}
& \multicolumn{1}{c}{Min}
& \multicolumn{1}{c}{Max} \\ \hline

$P_5$	&	1	&	0	&	1	&	1	\\ \hline
$P_6$	&	6.40E-02	&	1.40E-17	&	6.40E-02	&	6.40E-02	\\ \hline
$P_7$	&	1.00E-03	&	2.20E-19	&	1.00E-03	&	1.00E-03	\\ \hline
$P_8$	&	0.075	&	0.014	&	0.05	&	0.1	\\ \hline
$P_9$	&	1	&	0.0145	&	0.05	&	0.09	\\ \hline
$M_{6}$	&	0.585	&	0.287	&	0.0302	&	0.971	\\ \hline 
$M_{7}$	&	18.044	&	8.887	&	1	&	30.846	\\ \hline
$M_{8}$	&	0.484	&	0.0275	&	0.421	&	0.585	\\ \hline
$M_{9}$	&	0.0325	&	0.0012	&	0.027	&	0.0373	\\ \hline
$M_{10}$	&	0.0381	&	0.009	&	0.0209	&	0.089	\\ \hline
$M_{11}$	&	0.0525	&	0.0521	&	0.016	&	0.266	\\ \hline
$M_{12}$	&	0.9303	&	0.0332	&	0.8047	&	0.999	\\ \hline
$M_{13}$	&	508.306	&	268.197	&	11.184	&	1248.655	\\ \hline

\end{tabular}
    
\end{table}

In the context of a substantial dataset, it proves beneficial to ascertain its central tendency, often represented by a single value such as the mean, median, or mode. This central tendency provides an approximate average value, facilitating an understanding of the dataset's general characteristics. Referring to Table \ref{statanalysis}, it is evident that all numerical features exhibit a notably small standard deviation. This indicates that the data points for each feature are closely distributed around the mean, a sign of well-organized and reliable data. Additionally, to complement the numerical evaluation, we conducted a visual examination of the dataset. For instance, we inspected the distribution of one of the numerical features, namely the block interval feature ($P_8$). This analysis revealed a normal, or Gaussian, distribution, further validating the quality of the dataset. Furthermore, a meticulous inspection of the collected data did not identify any missing values. This absence of missing data implies that our dataset is complete and further contributes to the robustness of our subsequent ML analysis.

The correlation matrix between parameter-conditional features is helpful for understanding the data and examining feature relationships. This information can be used to verify projected performance. Table \ref{corr} presents the results of this analysis. We have found that the total number of blocks without transactions ($M_5$) is unrelated to the other features and can therefore be overlooked. However, the average block size ($M_6$) and the average number of transactions per block ($M_7$) have a strong positive association, demonstrating the power of the decision features.

\begin{table*}[ht]
    \centering
    \caption{Correlation matrix for the 22 features (9 parameters and 13 metrics) used in the experiments.}
    \label{corr}
    \footnotesize
\renewcommand{\arraystretch}{1.8}
    \resizebox{0.9\textwidth}{!}{%
    \begin{tabular}{cccccccccccccc |c cccccccccc}\cline{1-14} \cline{16-25}
{  }&	
{  $M_1$}	&	{  $M_2$	} &	{  $M_3$}	
&	{  $M_4$	} &	{  $M_5$}	&	{  $M_6$	}
&	{  $M_7$} &	{  $M_8$	} &	{  $M_9$}	
&	{  $M_{10}$}	&	{  $M_{11}$}	 &	{  $M_{12}$}	
&{  	$M_{13}$}	&&	\multicolumn{1}{c}{ {} } &	{  	$P_1$}		&	{  	$P_2$}		&	{  	$P_3$	}	&	{  $P_4$	}	&		{  $P_5$	}	&		{  $P_6$}		&		{  $P_7$	}	&		{  $P_8$	}	&	{  	$P_9$}		\\\cline{1-14} \cline{16-25}
           
{  $M_{1}$}		&	1		&	1		&	0.579		&	-0.099		&	0		&	0.25		&	0.25		&	-0.098		&	0.074		&	-0.93		&	-0.16		&	0.13		&	0.59	& \multicolumn{1}{c}{}&	{  $P_1$}		&		1		&		0		&		0		&		-0.03		&		0		&		0		&		0		&		-0.062		&		0		\\ \cline{1-14} \cline{16-25}

{  $M_{2}$	}	&	1		&	1		&	0.57		&	-0.09		&	0		&	0.25	&		0.25	&		-0.09	&		0.074	&		-0.93	&		-0.16		&	0.13		&	0.59	&\multicolumn{1}{c}{}&	{  $P_2$}		&		0		&		0		&		0		&		0		&		0		&		0		&		0		&		0		&		0		\\ \cline{1-14} \cline{16-25}

{  $M_{3}$}	&		0.57		&	0.57		&	1		&	0.30		&	0		&	0.90		&	0.9	&		0.38		&	-0.10		&	-0.50		&	0.42		&	0.58		&	0.99	&\multicolumn{1}{c}{}&	{  $P_3$	}	&		0		&		0		&		0		&		0		&		0		&		0		&		0		&		0		&		0		\\ \cline{1-14} \cline{16-25}

{  $M_{4}$}		&	-0.09		&	-0.09		&	0.30		&	1		&	0		&	0.42		&	0.43		&	0.41		&	-0.07		&	0.12		&	0.89		&	0.36		&	0.28	&\multicolumn{1}{c}{}&	{  $P_4$	}	&		-0.027		&		0		&		0		&		1		&		0		&		0		&		0		&		-0.11		&		0		\\ \cline{1-14} \cline{16-25}

{  $M_{5}$}		&	0		&	0		&	0		&	0		&	0		&	0		&	0		&	0		&	0		&	0		&	0		&	0		&	0	&\multicolumn{1}{c}{}&	{  $P_5$	}	&		0		&		0		&		0		&		0		&		0		&		0		&		0		&		0		&		0		\\ \cline{1-14} \cline{16-25}

{  $M_{6}$}	&		0.25	&		0.25	&		0.90	&		0.42	&		0		&	1		&	0.99		&	0.49		&	-0.08		&	-0.24		&	0.61		&	0.66		&	0.9	&\multicolumn{1}{c}{}&	{  $P_6$	}	&		0		&		0		&		0		&		0		&		0		&		0		&		0		&		0		&		0		\\ \cline{1-14} \cline{16-25}

{  $M_{7}$}		&	0.25		&	0.25		&	0.90		&	0.43		&	0		&	0.99		&	1		&	0.50		&	-0.12		&	-0.24		&	0.61		&	0.65	&	0.9	&\multicolumn{1}{c}{}&	{  $P_7$}		&		0		&		0		&		0		&		0		&		0		&		0		&		0		&		0		&		0		\\ \cline{1-14} \cline{16-25}

{  $M_{8}$}	&		-0.09		&	-0.09		&	0.38		&	0.41		&	0		&	0.49		&	0.50		&	1		&	-0.012		&	0.13		&	0.65	&	0.57		&	0.35	&\multicolumn{1}{c}{}&	{  $P_8$}		&		-0.062		&		0		&		0		&		-0.11		&		0		&		0		&		0		&		1		&		0		\\ \cline{1-14} \cline{16-25}

{  $M_{9}$}	&		0.07	&		0.07	&		-0.10	&		-0.07	&		0	&		-0.08		&	-0.12		&	-0.01		&	1		&	-0.12		&	-0.08		&	0.03		&	-0.1	&\multicolumn{1}{c}{}&	{  $P_9$	}	&		0		&		0		&		0		&		0		&		0		&		0		&		0		&		0		&		0		\\ \cline{1-14} \cline{16-25}

{  $M_{10}$}		&	-0.93		&	-0.93		&	-0.50		&	0.12		&	0		&	-0.24		&	-0.24		&	0.13		&	-0.12	&	1		&	0.18		&	-0.10		&	-0.52	&	\multicolumn{1}{c}{}	&	\multicolumn{1}{c}{}	&	\multicolumn{1}{c}{}	&	\multicolumn{1}{c}{}	&	\multicolumn{1}{c}{}	&\multicolumn{1}{c}{}		&	\multicolumn{1}{c}{}	&	\multicolumn{1}{c}{}	&	\multicolumn{1}{c}{}	&\multicolumn{1}{c}{}&	\multicolumn{1}{c}{}	\\ \cline{1-14} 

{  $M_{11}$	}&		-0.16		&	-0.16		&	0.426		&	0.89		&	0		&	0.61		&	0.61		&	0.65		&	-0.08		&	0.18		&	1		&	0.55		&	0.39	&	\multicolumn{1}{c}{}	&	\multicolumn{1}{c}{}	&	\multicolumn{1}{c}{}	&	\multicolumn{1}{c}{}	&	\multicolumn{1}{c}{}	& \multicolumn{1}{c}{}		&	\multicolumn{1}{c}{}	&	\multicolumn{1}{c}{}	&	\multicolumn{1}{c}{}	& \multicolumn{1}{c}{} &	\multicolumn{1}{c}{}	\\ \cline{1-14} 

{  $M_{12}$	}	&	0.13		&	0.13		&	0.58		&	0.36		&	0		&	0.66		&	0.65		&	0.57		&	0.03		&	-0.10		&	0.55		&	1		&	0.53	&	\multicolumn{1}{c}{}	&	\multicolumn{1}{c}{}	&	\multicolumn{1}{c}{}	&\multicolumn{1}{c}{}		&		\multicolumn{1}{c}{}&	\multicolumn{1}{c}{}	&	\multicolumn{1}{c}{}	&	\multicolumn{1}{c}{}	&	\multicolumn{1}{c}{}&\multicolumn{1}{c}{}	 &	\multicolumn{1}{c}{}	\\ \cline{1-14} 

{  $M_{13}$}	&	0.59		&	0.59		&	0.99		&	0.28		&	0		&	0.90		&	0.90		&	0.35		&	-0.10		&	-0.52		&	0.39		&	0.53		&	1	&	\multicolumn{1}{c}{}	&	\multicolumn{1}{c}{}	&	\multicolumn{1}{c}{}	&\multicolumn{1}{c}{}		&\multicolumn{1}{c}{}		&	\multicolumn{1}{c}{}	&\multicolumn{1}{c}{}		&	\multicolumn{1}{c}{}	&\multicolumn{1}{c}{}	&	\multicolumn{1}{c}{}&\multicolumn{1}{c}{}		\\ \cline{1-14} 

\end{tabular}
}
\end{table*}

\subsection{$k$NN Prediction results}
To prevent the issue of feature dominance, all numerical features are normalized. A normalized feature value $\widehat{v}_{u_{i},a_{j}}$ is obtained from its raw value $v_{u_{i},a_{j}}$ by 
\begin{equation*}
\widehat{v}_{u_{i},a_{j}}=\frac{v_{u_{i},a_{j}}-\underset{k}{\min }\left(
	v_{u_{k},a_{j}}\right) }{\underset{k}{\max }\left( v_{u_{k},a_{j}}\right) -%
	\underset{k}{\min }\left( v_{u_{k},a_{j}}\right) }\text{,}
\end{equation*}%
where $\underset{k}{\min }\left( v_{u_{k},a_{j}}\right) $ and $\underset{k}{%
	\max }\left( v_{u_{k},a_{j}}\right) $ are the minimum and maximum values of
feature $a_{j}$, considering all objects, respectively. This formula guarantees that 
$\widehat-{v}_{u_{i},a_{j}}\in \lbrack 0,1]$ for all $i$ and all $j$.

\begin{figure}[btp]
    \centering
    \scalebox{0.70}{ 
\begin{tikzpicture}[x=0.75pt,y=0.75pt,yscale=-1,xscale=1]

\draw  (40,274.74) -- (505,274.74)(86.5,51) -- (86.5,299.6) (498,269.74) -- (505,274.74) -- (498,279.74) (81.5,58) -- (86.5,51) -- (91.5,58)  ;
\draw [line width=1.2]     (105,61.6) -- (145,264.6) ;
\draw [line width=1.2]     (145,264.6) -- (166,194.6) ;
\draw [line width=1.2]     (166,194.6) -- (186,154.6) ;
\draw [line width=1.2]     (186,154.6) -- (206,261.6) ;
\draw [line width=1.2]     (206,261.6) -- (226,222.6) ;
\draw [line width=1.2]     (226,222.6) -- (249,224.6) ;
\draw [line width=1.2]     (269,153.6) -- (249,224.6) ;
\draw [line width=1.2]     (269,153.6) -- (288,143.6) ;
\draw [line width=1.2]     (288,143.6) -- (311,153.6) ;
\draw [line width=1.2]     (311,153.6) -- (330,136.6) ;
\draw [line width=1.2]     (330,136.6) -- (352,143.6) ;
\draw [line width=1.2]     (352,143.6) -- (370,128.6) ;
\draw [line width=1.2]     (370,128.6) -- (453,206.6) ;
\draw [line width=1.2]     (453,206.6) -- (477,202.6) ;
\draw [line width=1.2]   (477,202.6) -- (494,190.6) ;

\draw (110,283) node [anchor=north west][inner sep=0.75pt]  [font=\tiny] [align=left] {	 \ \ \ \ \ \ \ \ \ \ \ \ \ \ \ \ \ \ 2.5 \ \ \ \ \ \ \ \ \ \ \ \ \ \ \ \ \ \  \ \ \ \ \ \ \ \ \ \ \ \ \ 5.0 \ \ \ \ \ \ \ \ \ \ \ \ \ \ \ \ \ \ \ \ \ \ \ \ 7.5 \ \ \ \ \ \ \ \ \ \ \ \ \ \ \ \ \ \ \ \ \ \ 10.0\ \ \ \ \ \ \ \  \ \ \ \ \ \ \ \ \ \ \ \ \ \ \ 12.5\ \ \ \  \ \ \ \ \ \ \ \ \ \ \ \ \ \ \ \ \ \ 15.0\ \ \ \ \ \  \ \ \ \ \ \ \ \ \ \ \ \ \ \ \ \ 17.5 };
\draw (65,57) node [anchor=north west][inner sep=0.75pt]  [font=\tiny] [align=left] {85.0\\\\\\\\82.5\\\\\\\\80.0\\\\\\77.5\\\\\\\\75.0\\\\\\\\72.5\\\\\\\\70.0\\\\\\67.5};

\draw (275,310) node [anchor=north, font=\small] {Value of $k$};
\draw (40,190) node [anchor=west, font=\small, rotate=90] {RMSE};

\end{tikzpicture}
}
    \caption{RMSE as a function of the value of $k$. The figure shows that the optimal value of $k$ is 5, where the lowest RMSE is achieved.}
    \label{kvalues}
\end{figure}
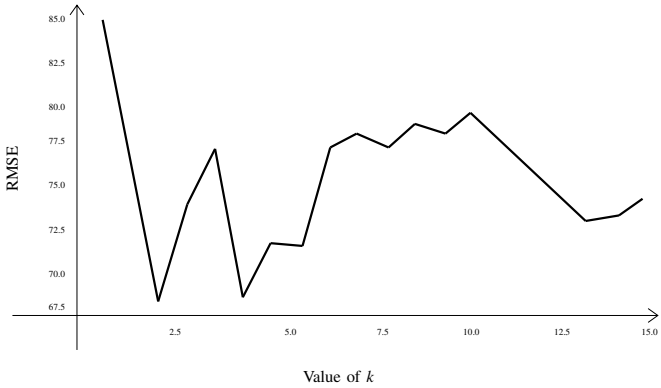

Our first test involves finding the best $k$ value for the $k$NN algorithm. To do so, we ran the model multiple times while changing the $k$ value and computing the root mean square error (RMSE). We then selected the $k$ value with the best RMSE. The RMSE is calculated as the standard deviation of the residuals, which represent the prediction errors. Residuals indicate how far data points are from the regression line, while RMSE indicates how spread out these residuals are. In other words, it shows how closely the data is clustered around the line of best fit. Root mean square error is often used to evaluate the results of experiments in climatology, forecasting, and regression analysis. RMSE is given by
\begin{equation*}
    RMSE = \sqrt{\frac{\sum_{i = 1} ^N ||y(i) - \hat{y}(i)||^2}{N}},
\end{equation*}

\noindent where $N$ represents the total count of data points, $y(i)$ denotes the $i$-th measurement in the dataset, and $\hat{y}(i)$ signifies the corresponding predictive estimation for the $i$-th observation. The result of this experiment is shown in Figure \ref{kvalues}. It is evident that a choice of $k=1$ leads to a very high RMSE. When $k$ is set to 5, the RMSE reduces significantly, approximating a value of 67.06. Any further increment in the value of $k$ results in a drastic drop in the RMSE. Consequently, it can be confidently inferred that $k=5$ is the optimal choice for this particular scenario, yielding the most favourable results.

$10$-fold cross-validation ensures solid results. Nine sub-datasets are for training and one for testing. Each object appears once in a test set and nine times in training sets. Results are averaged after $10$ separate tests. The aforesaid approach gave $k$NN 92\% accuracy, proving its applicability. We repeat the experiment with the SVM algorithm to be sure of the results. 89\% accuracy is below $k$NN.

Now, to test the model for an instance level, we feed the model with specific conditional features and predict some of the decision features. The prediction is made concerning both $k$NN and SVM. The results of this experiment are shown in Table \ref{comp}. The conditional feature values are copied from the first ten rows of the data. The interesting point is that the $k$NN prediction values are much closer to the target result than that of SVM. This ensures its correctness concerning blockchain performance prediction.

We used non-parametric approaches such as the Friedman methodology since the distribution of these data was uncertain \cite{Derrac2011}. For both its single and repeated testing options, this statistical investigation used MATLAB's Friedman's single and repeated test procedures. A Friedman function would construct a structure for the entire circumstance. This structure and a suitable post-hoc procedure will be used as input for a multi-comparison function. First, the Friedman test was performed with the null hypothesis of $k$NN and SVM. Friedman values for $k$NN, and SVM are 0.0015 and 0.0027, respectively. Because a significance level of $alpha=0.05$ was assumed, the null hypothesis is rejected in each case based on the p-values. As a result, we can now confirm that the two algorithms' accuracy is different. In other words, SVM is statistically different from $k$NN.

\begin{table}[tbh]
     \centering
     \caption{The classification accuracy of $k$NN and SVM for three different metrics over 10 different parameter configurations.}
     \label{comp}
     \setlength{\extrarowheight}{9pt}
     \setlength{\tabcolsep}{1pt}
    \resizebox{0.5\textwidth}{!}{%
    \begin{tabular}{cccccccccc|cccccc} 
    \hline 
   & \multicolumn{9}{c|}{Parameters ($P$)}  &  \multicolumn{6}{c}{Metrics ($M$)}     \\
       \hline

   $P_1$ & $P_2$ & $P_3$ & $P_4$ & $P_5$ & $P_6$ & $P_7$ & $P_8$ & $P_9$ && \multicolumn{2}{c}{$M_{11}$} &\multicolumn{2}{c}{$M_{12}$}&	\multicolumn{2}{c}{$M_{13}$} \\
   \cline{11-16}
   
    &\multicolumn{9}{c|}{}& $k$NN & SVM &	$k$NN &	SVM		
   &	$k$NN &	SVM		\\\hline

\multicolumn{1}{c}{13}&\multicolumn{1}{c}{1}&\multicolumn{1}{c}{raft}&\multicolumn{1}{c}{519}&\multicolumn{1}{c}{1}&\multicolumn{1}{c}{0.064}&	\multicolumn{1}{c}{0.001}&\multicolumn{1}{c}{0.083}&\multicolumn{1}{c}{1}&&\multicolumn{1}{c}{0.033} &\multicolumn{1}{c}{0.024}&\multicolumn{1}{c}{\textbf{0.912}} &\multicolumn{1}{c}{0.99}&\multicolumn{1}{c}{\textbf{569.026}}&\multicolumn{1}{c}{559.99} \\ \hline

\multicolumn{1}{c}{6}&\multicolumn{1}{c}{1}&\multicolumn{1}{c}{raft}&\multicolumn{1}{c}{682}&\multicolumn{1}{c}{1}&\multicolumn{1}{c}{0.064}&\multicolumn{1}{c}{0.001}&\multicolumn{1}{c}{0.069}&	\multicolumn{1}{c}{1}&&\multicolumn{1}{c}{0.035}&\multicolumn{1}{c}{0.045}&\multicolumn{1}{c}{\textbf{0.92}}&\multicolumn{1}{c}{0.89}&\multicolumn{1}{c}{\textbf{737.72}}&\multicolumn{1}{c}{744.66} \\ \hline

\multicolumn{1}{c}{9}&\multicolumn{1}{c}{1}&\multicolumn{1}{c}{raft}&\multicolumn{1}{c}{66}&\multicolumn{1}{c}{1}&\multicolumn{1}{c}{0.064}&	\multicolumn{1}{c}{0.001}&	\multicolumn{1}{c}{0.070}&	\multicolumn{1}{c}{1}&&\multicolumn{1}{c}{0.022}& \multicolumn{1}{c}{0.055}&\multicolumn{1}{c}{	\textbf{0.91}} &\multicolumn{1}{c}{0.84}	&\multicolumn{1}{c}{\textbf{72.35}}&\multicolumn{1}{c}{72.44} \\ \hline

\multicolumn{1}{c}{9}&\multicolumn{1}{c}{1}&\multicolumn{1}{c}{raft}&\multicolumn{1}{c}{450}&\multicolumn{1}{c}{1}&\multicolumn{1}{c}{0.064}&	\multicolumn{1}{c}{0.001}&	\multicolumn{1}{c}{0.058}&	\multicolumn{1}{c}{1}&&\multicolumn{1}{c}{0.020}&\multicolumn{1}{c}{0.029}&\multicolumn{1}{c}{\textbf{0.93}}&\multicolumn{1}{c}{0.83}&\multicolumn{1}{c}{\textbf{480.88}}&\multicolumn{1}{c}{489.81} \\ \hline

\multicolumn{1}{c}{15}&\multicolumn{1}{c}{1}&\multicolumn{1}{c}{raft}&\multicolumn{1}{c}{893}&\multicolumn{1}{c}{1}&\multicolumn{1}{c}{0.064}&\multicolumn{1}{c}{0.001}&\multicolumn{1}{c}{0.072}&	\multicolumn{1}{c}{1}&&\multicolumn{1}{c}{0.12}&\multicolumn{1}{c}{0.19}&\multicolumn{1}{c}{\textbf{0.99}}&\multicolumn{1}{c}{0.74}&\multicolumn{1}{c}{\textbf{754.931}}&\multicolumn{1}{c}{759.899} \\\hline

\multicolumn{1}{c}{9}&\multicolumn{1}{c}{1}&\multicolumn{1}{c}{raft}&\multicolumn{1}{c}{440}&\multicolumn{1}{c}{1}&\multicolumn{1}{c}{0.064}&\multicolumn{1}{c}{0.001}&\multicolumn{1}{c}{0.069}&	\multicolumn{1}{c}{1}&&\multicolumn{1}{c}{0.026}&\multicolumn{1}{c}{0.031}&\multicolumn{1}{c}{\textbf{0.940}}&\multicolumn{1}{c}{0.830}&\multicolumn{1}{c}{\textbf{467.88}}&\multicolumn{1}{c}{476.35} \\ \hline

\multicolumn{1}{c}{6}&\multicolumn{1}{c}{1}&\multicolumn{1}{c}{raft}&\multicolumn{1}{c}{965}&\multicolumn{1}{c}{1}&\multicolumn{1}{c}{0.064}&\multicolumn{1}{c}{0.001}&	\multicolumn{1}{c}{0.065}&	\multicolumn{1}{c}{1}&&\multicolumn{1}{c}{0.1077}&\multicolumn{1}{c}{0.098}&\multicolumn{1}{c}{\textbf{0.98}}&\multicolumn{1}{c}{0.98}&\multicolumn{1}{c}{\textbf{982.21}}&\multicolumn{1}{c}{977.23} \\ \hline

\multicolumn{1}{c}{7}&\multicolumn{1}{c}{1}&\multicolumn{1}{c}{raft}&\multicolumn{1}{c}{17}&\multicolumn{1}{c}{1}&\multicolumn{1}{c}{0.064}&\multicolumn{1}{c}{0.001}&\multicolumn{1}{c}{0.095}&	\multicolumn{1}{c}{1}&&\multicolumn{1}{c}{0.033}&\multicolumn{1}{c}{0.055}&\multicolumn{1}{c}{\textbf{0.91}}&\multicolumn{1}{c}{0.97}&\multicolumn{1}{c}{\textbf{18.68}}&\multicolumn{1}{c}{19.67}\\\hline
\end{tabular}%
}
\end{table}

We proceeded on to the second step, which was dubbed ``post hoc," once we realised that the accuracy of the two algorithms was not the same. During this phase, we conducted four Friedman tests for each of the four different situations. Because repeated testing leads to an increased risk of making a Type I error—that is, incorrectly concluding that a null hypothesis should be rejected when in fact it should be accepted—we were forced to use one of the post hoc methods at our disposal in order to find a solution to this issue. We used two, Fisher's least significant difference approach \cite{Ulrich2006} and Tukey's honest significant difference criteria \cite{Nanda2021}. Both of these are described in the references. The p-values that were determined as a consequence of all of these tests may be seen in Table \ref{posthoc}, which also contains a plethora of other information. With a closer look, we can notice that the majority of the values are less than 0.05 which confirms the statistical difference between the algorithms.

\begin{table}[ht]
	\centering
	\caption{Post hoc $p$-values resulting from Friedman tests of  $K$NN for four parameters.}
	\label{posthoc}
	\par
	\renewcommand{\arraystretch}{1} \centering
	\par
	{\footnotesize \ }
	\par
	{\footnotesize \ 
\setlength{\tabcolsep}{3pt}
    \linespread{0.4}
    \begin{tabular}{cccccccc}
			\toprule 
			&  & Correction & $P_5$ & $P_6$ & $P_7$ & $P_8$ & $P_9$ \\
			\midrule 
			& \multirow{2}{*}{Features} & Fisher & 0.0001 & 0.0007 & 0.0575 & 0.1372 & 0.0166  \\
			&   & Turkey & 0.0028 & 0.0126 & 0.4808 & 0.7532 & 0.2007  \\\hline
			
		\end{tabular}
	}
\end{table}

\subsection{ISO validation results}
This section looks at how well the ISO algorithm works to find parameters in a blockchain. We compare ISO's performance to that of five other competitor algorithms to make the investigation meaningful. The five algorithms we compare, namely, PSO, HHO, GWO, ABC, ACO, and SO, are very recent.
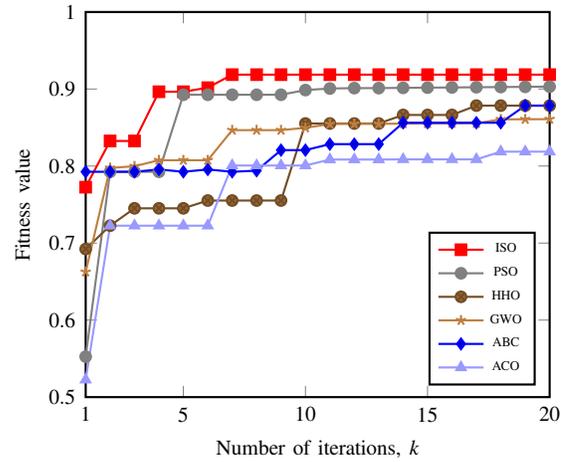
\begin{figure}[htp]
    \centering
    \begin{tikzpicture} \begin{axis}[xmin=1,
		xmax=20,xtick={1,5,10,15,20},ymin=0.5,ymax=1, xlabel={Number of iterations,
			$k$}, ylabel={{Fitness value}},legend pos=south east, style={ thick, font =
			\footnotesize}, scale = 0.9,legend style = {font = \tiny}] \addplot plot [color =
		red, mark = square*, mark options = {fill = red!150}] coordinates { ( 1 ,
			0.77255639 ) ( 2 , 0.83255639 ) ( 3 , 0.83255639 ) ( 4 , 0.89646616 ) ( 5 ,
			0.89646616 ) ( 6 , 0.90172932 ) ( 7 , 0.91864661 ) ( 8 , 0.91864661 ) ( 9 ,
			0.91864661 ) ( 10 , 0.91864661 ) ( 11 , 0.91864661 ) ( 12 , 0.91864661 ) (
			13 , 0.91864661 ) ( 14 , 0.91864661 ) ( 15 , 0.91864661 ) ( 16 , 0.91864661
			) ( 17 , 0.91864661 ) ( 18 , 0.91864661 ) ( 19 , 0.91864661 ) ( 20 ,
			0.91864661 ) }; \addplot plot [color = gray,mark = *, mark options = {fill =
			gray}, legend style = {font = \tiny}] coordinates { ( 1 , 0.55255639 ) ( 2 ,
			0.79255639 ) ( 3 , 0.79255639 ) ( 4 , 0.79255639 ) ( 5 , 0.89255639 ) ( 6 ,
			0.89255639 ) ( 7 , 0.89255639 ) ( 8 , 0.89255639 ) ( 9 , 0.89255639 ) ( 10 ,
			0.89864661 ) ( 11 , 0.90064661 ) ( 12 , 0.90116616 ) ( 13 , 0.90116616 ) (
			14 , 0.90151253 ) ( 15 , 0.90177231 ) ( 16 , 0.90203208 ) ( 17 , 0.90229186
			) ( 18 , 0.90255164 ) ( 19 , 0.90281141 ) ( 20 , 0.90307119 ) }; \addplot
		plot coordinates { ( 1 , 0.692255639 ) ( 2 , 0.72255639 ) ( 3 , 0.74511378 )
			( 4 , 0.74511378 ) ( 5 , 0.74511378 ) ( 6 , 0.7552138 ) ( 7 , 0.7552138 ) (
			8 , 0.7552138 ) ( 9 , 0.7552138 ) ( 10 , 0.85521379 ) ( 11 , 0.85521379 ) (
			12 , 0.85521379 ) ( 13 , 0.85521379 ) ( 14 , 0.8663249 ) ( 15 , 0.8663249 )
			( 16 , 0.8663249 ) ( 17 , 0.87844612 ) ( 18 , 0.87844612 ) ( 19 , 0.87844612
			) ( 20 , 0.87844612 ) }; \addplot plot [color = brown] coordinates { ( 1 ,
			0.66255639 ) ( 2 , 0.79755639 ) ( 3 , 0.799755639 ) ( 4 , 0.80755639 ) ( 5 ,
			0.80755639 ) ( 6 , 0.80755639 ) ( 7 , 0.84661238 ) ( 8 , 0.84661238 ) ( 9 ,
			0.84661238 ) ( 10 , 0.84965749 ) ( 11 , 0.85454624 ) ( 12 , 0.85480602 ) (
			13 , 0.85480602 ) ( 14 , 0.85507957 ) ( 15 , 0.85520946 ) ( 16 , 0.85533935
			) ( 17 , 0.85546924 ) ( 18 , 0.86059912 ) ( 19 , 0.86072901 ) ( 20 ,
			0.8608589 ) }; \addplot plot coordinates { ( 1 , 0.79255639 ) ( 2 ,
			0.79255639 ) ( 3 , 0.79255639 ) ( 4 , 0.7955639 ) ( 5 , 0.79255639 ) ( 6 ,
			0.7955639 ) ( 7 , 0.79255639 ) ( 8 , 0.79400168 ) ( 9 , 0.82066838 ) ( 10 ,
			0.82066838 ) ( 11 , 0.82844588 ) ( 12 , 0.82844588 ) ( 13 , 0.82844588 ) (
			14 , 0.8562239 ) ( 15 , 0.8562239 ) ( 16 , 0.8562239 ) ( 17 , 0.8562239 ) (
			18 , 0.8562239 ) ( 19 , 0.87844612 ) ( 20 , 0.87844612 ) }; \addplot plot
		[solid,color = blue!40,mark = triangle*, mark options = {fill = blue!40}]
		coordinates { ( 1 , 0.52255639 ) ( 2 , 0.72255639 ) ( 3 , 0.72255639 ) ( 4 ,
			0.72255639 ) ( 5 , 0.72255639 ) ( 6 , 0.72255639 ) ( 7 , 0.80066838 ) ( 8 ,
			0.80066838 ) ( 9 , 0.80066838 ) ( 10 , 0.80066838 ) ( 11 , 0.80844588 ) ( 12
			, 0.80844588 ) ( 13 , 0.80844588 ) ( 14 , 0.80864661 ) ( 15 , 0.80864661 ) (
			16 , 0.80864661 ) ( 17 , 0.80864661 ) ( 18 , 0.81864661 ) ( 19 , 0.81864661
			) ( 20 , 0.81864661 ) };
		\legend{ISO\\PSO\\HHO\\GWO\\ABC\\ACO\\} \end{axis}
		\end{tikzpicture}
    \caption{Fitness value compared to the number of iterations: A higher fitness value indicates quicker convergence of the algorithm.}
    \label{performance-comparison}
\end{figure}

Here, each algorithm searches for the ideal configuration based on a metric input value. To verify, the $k$NN regressor receives the parameter vector. The method is more reliable the closer the original value is to the anticipated one. We used 20 salps for 50 iterations in this experiment with three leaders. Table \ref{validation} shows that ISO (last row) won this experiment. In the last row, $M {13} = 1100$, ISO produced parameter vector has 83\% accuracy, while the best competitor, the classic salp technique, has 81 accuracy.
\begin{table*}[htb]
    \centering
    \caption{The fitness value achieved by ISO and six competitors. Clearly, ISO (the last row) comes out as a clear winner.}
    \label{validation}
\resizebox{2\columnwidth}{!}{%
\normalsize
\renewcommand{\arraystretch}{1.8}
    \begin{tabular}{lccccccccccccc}\toprule 
     \multicolumn{13}{c}{Metrics ($M$)} \\
    \cmidrule(lr){2-14} 
        \multicolumn{1}{c}{Algo.} &	\multicolumn{1}{c}{$M_{1} = 33$}		&\multicolumn{1}{c}{$M_{2} = 29$}		&	\multicolumn{1}{c}{$M_{3} = 1102$}		&\multicolumn{1}{c}{$M_{4} = 50$}		&\multicolumn{1}{c}{$M_{5} = 0.5$}		&\multicolumn{1}{c}{$M_{6} = 0.7$}		&\multicolumn{1}{c}{$M_{7} = 25$}		&\multicolumn{1}{c}{$M_{8} = 0.2$}		&\multicolumn{1}{c}{$M_{9} = 0.02$}		&\multicolumn{1}{c}{$M_{10} = 0.07$}		&	\multicolumn{1}{c}{$M_{11} = 0.25$}		&	\multicolumn{1}{c}{$M_{12} = 0.8$}		&	\multicolumn{1}{c}{$M_{13}=1100$}		\\\hline
	PSO		&		21		&		25		&		559		&		22		&		0.4		&		0.5		&		21		&		0.2		&		0.018		&		0.06		&		0.15		&		0.4		&		752		\\ \hline
	HHO		&		25		&		21		&		687		&		29		&		0.5		&		0.6		&		21		&		0.18		&		0.011		&		0.05		&		0.19		&		0.5		&		897		\\ \hline
	GWO		&		26		&		20		&		714		&		25		&		0.4		&		0.7		&		22		&		0.15		&		0.019		&		0.06		&		0.18		&		0.5		&		774		\\ \hline
	ABC		&		26		&		27		&		752		&		26		&		0.3		&		0.5		&		24		&		0.2		&		0.025		&		0.06		&		0.19		&		0.8		&		687		\\ \hline
	ACO 		&		26		&		28		&		777		&		29		&		0.4		&		0.5		&		20		&		0.19		&		0.23		&		0.05		&		0.19		&		0.7		&		744		\\ \hline
	SO		&		29		&		26		&		798		&		29		&		0.4		&		0.7		&		22		&		0.2		&		0.03		&		0.05		&		0.22		&		0.8		&		899		\\ \hline
	ISO		&		\textbf{31}		&		\textbf{29}		&		\textbf{912}		&		\textbf{43}		&		\textbf{0.5}		&		\textbf{0.7}		&		\textbf{24}		&		\textbf{0.2}		&		\textbf{0.021}		&	\textbf{	0.07}		&		\textbf{0.23}		&		\textbf{0.8}		&		\textbf{915}		\\\hline

    \end{tabular}
    }
\end{table*}

The development of the fitness value across iterations serves as another comparison test. Figure \ref{performance-comparison} shows this trend. The ISO curve is generally superior to all other curves. This suggests that ISO consistently outperforms other standards, regardless of the statistic. As a result, the evolution paints a clear picture of the algorithm's conduct from the beginning to the finish of the assignment. 

\begin{table}[tbh]
    \centering
    \caption{The predicted parameters form the 6 algorithm for $m_{13} = 1100$. Clearly, ISO reached a parameter vector achieving the closest value.}
    \label{parameterResults}
    \footnotesize
\renewcommand{\arraystretch}{1.3}
    \setlength{\extrarowheight}{5pt}
    \setlength{\tabcolsep}{1pt}
    \resizebox{0.5\textwidth}{!}{%
    \begin{tabular}{cccccccccc c}\toprule
    \multirow{2}{*}{Algo.} & \multicolumn{9}{c}{parameters ($P$)} & \multirow{2}{*}{Achieved value} \\
    \cmidrule(lr){2-10} 
     &$P_1$	&	$P_2$	&	$P_3$	&	$P_4$	&	$P_5$	&	$P_6$	&	$P_7$	&	$P_8$	&	$P_9$&\\\hline
    \multicolumn{1}{c}{PSO}&\multicolumn{1}{c}{6}&\multicolumn{1}{c}{ 1}&\multicolumn{1}{c}{ 1}&\multicolumn{1}{c}{ 654.670}&\multicolumn{1}{c}{1}&\multicolumn{1}{c}{ 0.368}&\multicolumn{1}{c}{ 0.557}& \multicolumn{1}{c}{0.0796}&\multicolumn{1}{c}{1}&\multicolumn{1}{c}{681.390} \\ \hline
    
    \multicolumn{1}{c}{HHO}&\multicolumn{1}{c}{6}&\multicolumn{1}{c}{1}&\multicolumn{1}{c}{1}& \multicolumn{1}{c}{1087.262}&\multicolumn{1}{c}{1}&\multicolumn{1}{c}{0.264}&\multicolumn{1}{c}{ 0.509}&\multicolumn{1}{c}{ 0.0752}&\multicolumn{1}{c}{1}&\multicolumn{1}{c}{806.143} \\ \hline
    
    \multicolumn{1}{c}{GWO}&\multicolumn{1}{c}{6}&\multicolumn{1}{c}{1}&\multicolumn{1}{c}{1}&\multicolumn{1}{c}{733.183}&\multicolumn{1}{c}{1}&\multicolumn{1}{c}{0.428} &\multicolumn{1}{c}{0.441}&\multicolumn{1}{c}{0.0787}&\multicolumn{1}{c}{1}&\multicolumn{1}{c}{777.512}\\ \hline
    
    \multicolumn{1}{c}{ABC}&\multicolumn{1}{c}{8}&\multicolumn{1}{c}{1}&\multicolumn{1}{c}{1}&\multicolumn{1}{c}{796.841}&\multicolumn{1}{c}{1}&\multicolumn{1}{c}{0.393}&\multicolumn{1}{c}{0.501}&\multicolumn{1}{c}{0.068}&\multicolumn{1}{c}{1}&\multicolumn{1}{c}{739.642} \\ \hline
    
    \multicolumn{1}{c}{SO}&\multicolumn{1}{c}{6}&\multicolumn{1}{c}{1}&\multicolumn{1}{c}{1}&\multicolumn{1}{c}{750.958}&\multicolumn{1}{c}{1}&\multicolumn{1}{c}{0.347}&\multicolumn{1}{c}{0.401}&\multicolumn{1}{c}{ 0.059}&\multicolumn{1}{c}{1}&\multicolumn{1}{c}{766.865}\\\hline
    
    \multicolumn{1}{c}{ISO}&\multicolumn{1}{c}{6}&\multicolumn{1}{c}{1}&\multicolumn{1}{c}{1}&\multicolumn{1}{c}{ 919.76}&\multicolumn{1}{c}{1}&\multicolumn{1}{c}{0.297}&\multicolumn{1}{c}{ 0.567}&\multicolumn{1}{c}{0.064}&\multicolumn{1}{c}{1}&\multicolumn{1}{c}{ \textbf{823.940}}\\\hline
         \end{tabular}
    }
\end{table}
One final experiment is to look at the predicted parameter vector by ISO and its five competitors. In this experiment, we are attempting to identify the parameter vector that will result in $M_{13} = 1100$. The results of this experiment are shown in the table \ref{parameterResults}. Noting that the last column (achieved value) corresponds to the value of $M_{13}$ in the resulting parameter vector, we can notice that ISO has the best-achieved value which is much closer to 1100 than any other competitor. 

\section{Conclusions}
\label{conc}
The advent of blockchain technology has initiated a paradigm shift across numerous sectors due to its inherent potential and advanced capabilities. Nevertheless, the intricate and decentralized characteristics of blockchain's underlying infrastructure introduce challenges in assessing the performance of blockchain-based applications. Thus, the necessity for a dependable modeling methodology becomes paramount to aid the creation and performance evaluation of such applications. Historically, research has predominantly focused on simulation-based solutions to evaluate blockchain application performance, while the exploration of machine learning (ML) model-based techniques remains comparatively scant in this context. This study sought to bridge this gap by proposing two innovative ML-based techniques.

The first approach integrated a $k$ nearest neighbour ($k$NN) and a support vector machine (SVM) to predict blockchain performance by leveraging predefined configuration parameters. The second method utilized salp swarm optimization (SO), an ML model, to identify the most advantageous blockchain configurations for achieving the desired performance benchmarks. To further refine the efficacy of SO, we incorporated rough set theory, thus formulating an Improved Swarm Optimization (ISO) model. The ISO model displayed superior capabilities in generating accurate recommendations for optimal parameter configurations amidst uncertainties. Upon comparative statistical evaluation, our proposed models exhibited a competitive advantage. Specifically, the $k$NN model outperformed the SVM by a margin of 5\%, while the ISO model demonstrated a 4\% reduction in accuracy deviation relative to the standard SO model. These encouraging results underscore the potential of our proposed methodology in addressing the inherent challenges associated with evaluating the performance of blockchain-based applications. Moreover, they underline the contribution of our work to the advancement and performance evaluation of blockchain-based applications.

Notably, the utility of our models is not confined to specific algorithms, thereby enhancing their adaptability. Future research directions include investigating the scalability of our proposed methodology and its applicability to larger, more complex blockchain-based applications. Additionally, the exploration of a broader range of recent algorithms beyond SVM and $k$NN presents an exciting avenue for future studies.

\section*{Acknowledgements}
 This work is funded in part by the EPSRC, under grant number EP/V042017/1. Scalable Circular Supply Chains for the Built Environment.

\bibliographystyle{IEEEtran}
\bibliography{references}
\end{document}